# TSET: Token based Secure Electronic Transaction

Rajdeep Borgohain, Moirangthem Tiken Singh
Department of Computer Science and Engineering
Dibrugarh University Institute of Engineering and Technology
Dibrugarh, India
rajdeepgohain@gmail.com,tiken09@gmail.com

Chandrakant Sakharwade
DKOP Labs Pvt. Ltd
NOIDA 201301, India
chandrakant.sakharwade@gmail.com

Sugata Sanyal
School of Technology and Computer Science,
Tata Institute of Fundamental Research,
Mumbai, India
sanyals@gmail.com

*Abstract* — **Security and trust are the most important factors in online transaction, this paper introduces TSET a Token based Secure Electronic Transaction which is an improvement over the existing SET, Secure Electronic Transaction protocol. We take the concept of tokens in the TSET protocol to provide end to end security. It also provides trust evaluation mechanism so that trustworthiness of the merchants can be known by customers before being involved in the transaction. Moreover, we also propose a grading mechanism so that quality of service in the transactions improves.**

*Keywords- SET Protocol; Token; Privacy; End to End Security,TTP.*

## I.  INTRODUCTION

Mobile payment is the transaction of fiscal values by means of mobile phones or other handheld devices. According to one of the Gartner's report (Christy Pettey, 2011) the total mobile users in the world will reach 7.4 billion by 2015. With such a large number of people using mobile devices, it would be increasingly used not only for communication but also as a means of monetary transactions (Melissa Soo Ding and Chandan R. Unnithan, 2002).As mobile phones have become more and more powerful with multiple features, people would rather like to have their monetary transaction done with a mobile device rather than carrying currencies and notes in their pocket.

Though there are many existing mobile payment protocols, one of the most widely accepted mobile payment protocol is the Secure Electronic Transaction protocol. (Zhang Boping and Shang Shiyu, 2009)  Though SET has been accepted by many companies as the standard security protocol for online transactions, SET still has issues which need to be addressed (Noureddine Boudriga, 2009). SET does not provide a way for the customers to know the trustworthiness of the party they are dealing with.

 (Xun-yi Ren et al., 2011).This lack of trust is one of the prime reason people abstain from participating in online transactions and this has been a major hurdle for e-commerce. If there was a mechanism to know a priori the trustworthiness of the party the customers are dealing with, people would be more open to e-commerce. SET also does not guarantee the quality of products that will be available to the customers after the transaction, i.e. if the customer is not satisfied with the quality of the product after the transaction then SET does not provide any mechanism by which the merchant becomes liable to provide a replacement or refund the amount of the product. Moreover, SET protocol does not provide any mechanism for end to end security (Ayu Tiwari et al., 2007). The request for transaction can be compromised by any agency at any point in the transaction process (Rangarajan A. Vasudevan and Sugata Sanyal, 2004) and lot more amount of money may be transacted than allowed by the customer without the knowledge of the customer.

In this paper, we propose a method which enables people to know in advance the trustworthiness of the party customers are dealing with, provide a mechanism by which the customers would receive intended goods and provide end to end security of the transaction. To achieve the end to end security we introduce the concept of tokens which are generated by the customer's bank based on which the transaction would be carried out. Any tampering in the token would indicate that the amount value in the transaction has been compromised and the transaction would not be allowed.

The rest of the paper is organized in the following way: Section 2 gives an overview of the SET protocol. In Section 3 we look at the disadvantages of the SET protocol. In Section 4 we introduce and discuss the TSET protocol. Section 5 gives an analysis of the TSET protocol and we finally conclude the paper in Section 6.

## II. OVERVIEW OF SET PROTOCOL

The SET protocol is a security specification introduced by VISA and MasterCard for secure transaction over the internet. The main aim of the SET protocol is to ensure confidentiality of information. Secondly, it ensures the integrity of all the data that are transmitted during the transaction process. Finally, the SET protocol provides authentication that both the customer and the merchant are legitimate (Yang Li and Yun Wang, 2001). Both the customer and the merchant are provided with digital certificates that authenticate their legitimacy to make transaction over the network. The SET protocol basically involves the following entities: a Customer (Cardholder), Customer Bank (Issuer), Merchant, and Merchant Bank (Acquirer). Before participating in the transaction both the customer and the merchant must obtain a digital certificate from a Certifying Authority.

The steps involved in the SET protocol are:

1. The customer browses the website of the merchant and chooses the product.

2. The merchant returns a form containing the list of items along with total price and order number. A copy of digital certificate is also sent for the authentication of the merchant.

3. The customer sends the dual signature order information and the payment information along with customer digital certificate. The digital certificate is to validate the customer's authenticity. The order information confirms that the customer will make the purchase, whereas the payment information is encrypted by the public key of the payment gateway which cannot be read by the merchant.

4. The merchant forwards the payment information to the merchant bank.

5. The merchant bank then forwards the information to the Customer Bank for authorization and payment.

6. The Customer Bank sends authorization to the merchant bank and merchant bank sends the authorization to the merchant.

7. The merchant completes the order and sends it to the customer.

8. The merchant captures the transaction from their bank.

9. The Customer Bank sends a notification to the Customer that the payment has been processed.

The data model of the SET protocol is given in Figure 1.

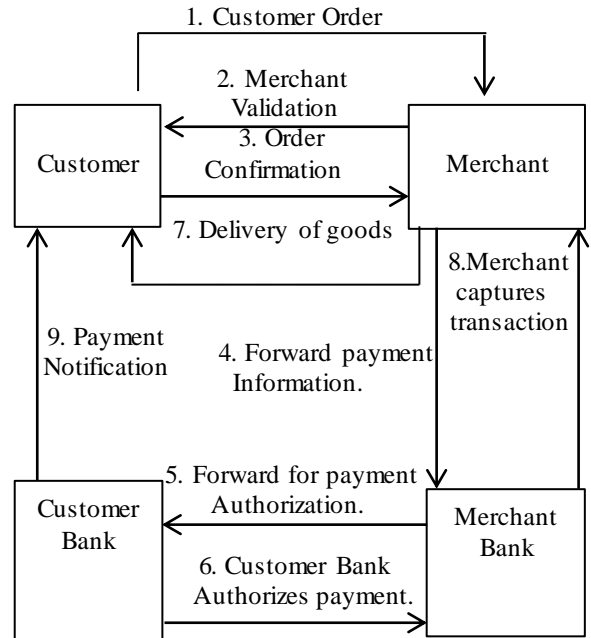

Figure 1 The SET Protocol

The SET protocol was succeeded by 3D SET *(VISAEU, 3D SET)*. Visa and MasterCard introduced 3D SET in 1999, to facilitate flexibility and portability for customers. In 3D SET, the transaction logs are stored in the banks, as banks were deemed trusted entities by the customers (Jing Jang Hwang et al., 2003)

## III. LIMITATIONS OF SET PROTOCOL

The limitations of SET protocol are:
1. In the SET protocol there is no way in which the customer knows the trustworthiness of the merchant he is dealing with. (Xun-yi Ren et al.,2011) The customers remain ignorant whether to trust the merchant with the deal or not. This is the main reason many people remains skeptical about e-commerce, based on SET protocol.

2. The SET protocol does not provide any means by which the customer is assured that the goods that will be sent to him will be of the desired quality. If the products are not as per liking of the customer, the customer must be able to get a replacement or get a refund. This is not guaranteed by the SET protocol.

3. The SET protocol does not guarantee end to end security. During the transaction process the network

may be hacked by any agencies at any point. If this happens the customer will end up paying much more than he intended to do without his knowledge. Moreover, in the traditional flow of transaction, there is fear of modification of balance by merchants. (Sugata Sanyal et al., 2010; Vipul Goyal et al., 2005).

4. In SET protocol, the privacy of the customer is compromised (Tan Soo Fun et al., 2008). Even in 3D SET the information regarding the payment and the order lies with the bank entities. (Jing Jang Hwang. et al., 2003). The private information of the customers and the merchants can be accessed by the banks which could be misused by any third party who could get access to this information.

## IV. THE TSET PROTOCOL

The proposed protocol TSET, addresses the issues relating to trustworthiness of the merchants, ensuring customer satisfaction of the goods and end to end security. The TSET protocol involves the following entities, Customer, Merchant, Customer Bank, Merchant Bank and a Trusted Third Party (TTP). Usually, SSL(Secure Socket Layer) protocol is used by the TTP (Xu Yong and Liu Jindi, 2010). In this model, the Trusted Third Party works as a moderator between all the entities involved. The TTP stores the transaction log of the deals. In case of any disputes, the transaction log stored with the TTP is used to resolve the issues. It provides an undeniable proof of the transaction between the parties and as such issues like non repudiation cannot be raised. The TTP is also responsible for keeping track of the trust factor of the merchants. The trust factor of the merchant is stored with the TTP which the customers can access to analyze the trustworthiness of the merchant before getting involved with the transaction. Finally, the TTP also acts as an arbitrary body in case of any disputes. The symbols used in the protocol are given in Table 1:

| Symbol | Meaning |
|---|---|
| C | Customer |
| M | Merchant |
| CB | Customer Bank |
| MB | Merchant Bank |
| $DCert_X$ | Digital Certificate of entity X |
| AM | Amount |
| TS | Timestamp |
| $TF_X$ | Trust Factor of entity X |
| OI | Order Information |
| TKN | Token |
| $TKN_{req}$ | Token Request |
| $PK_X$ | Public Key of entity X |
| $SK_X$ | Secret Key of entity X |
| $TID_X$ | Transaction ID by entity X |

Table1. Symbols used in the protocol

### A. Calculation of the Trust Factor:

The most important factor in the online transaction is the issue of trust. Even though people have the convenience of making online transactions from home and having products delivered at their doorsteps, people are still hesitant to indulge in online transaction activities. People are unsure of the other party which they deal with. If there is a mechanism which informs the customer, prior to the online transaction, how trustworthy the merchants are, people would be more open towards electronic commerce.

For calculating the trust factor of a merchant we propose a simple technique. The trust factor of a merchant involved in the transaction would remain with the TTP. At any point of time when the customer is about to involve in an online deal, he can log on into the website of the TTP and analyze the trust factor of the merchant. If the customer finds the trust factor of the merchant satisfactory he can proceed with the transaction and if he is not convinced with the trust factor of the merchant he can back away from the transaction. To trust factor can be calculated by the following formula,

$$TF_M = 100 - TV_M,$$

where $TF_M$ = Trust Factor of Merchant.
$TV_M$ = Trust Value of Merchant.

The trust value of a merchant is decided by the total number of transaction the merchant is involved in and the total number of transactions where there had been a customer complaint and initial products were rejected. The trust value is calculated as,

$$TV_M = \frac{Number\ of\ times\ rejected\ by\ the\ customer}{Total\ number\ of\ times\ involved\ in\ transaction} \times 100$$

When a merchant gets refund or replacement request from the same customer for the same product more than once the trust value is calculated as $(TV_M)^2$, so that the customer does not have to go through the same ordeal again and again.

The trust factor is divided into different grades given in Table 2.

| $TF_M$ | GRADE | $TF_M$ | GRADE |
|---|---|---|---|
| 100-90 | A1 | 49-40 | C2 |
| 89-80 | A2 | 39-30 | D1 |
| 79-70 | B1 | 29-20 | D2 |
| 69-60 | B2 | 19-10 | E1 |
| 59-50 | C1 | 9-0 | E2 |

Table 2 Grades distribution

Now, during the transaction the customer can view the trust factor of the merchant and decide himself whether he wants to participate in the transaction or not. For example, a certain merchant $M_1$ has a total of 1000 transactions and among them 25 of the transaction had replacement or refund of goods. So the trust value of the merchant $M_1$ becomes:

$$TV_{M1} = \frac{25}{1000} \times 100$$

$$= 2.5$$

So, the total trust factor $TF_{M1}$ would be 97.5, which would assign a grade A1 to the merchant $M_1$.

On the other hand if a merchant $M_2$ has 300 refunds out of a total 1000 transactions, the trust value of merchant $M_2$ would become:

$$TV_{M2} = \frac{300}{1000} \times 100$$

$$= 30$$

So, the total trust factor $TF_{M2}$ would be 70, which would assign a grade B1 to the merchant $M_2$.

Given a choice between the merchants, the customer would always go for merchant with the higher trust factor. This would give the customer a greater sense of trust to get involved in online transaction. Moreover, the merchants would always strive to provide highest quality of service to the customers so that their trust factor always remain as high as possible.

### B. Format of the Token:

For every transaction the Customer Bank generates a token which contains the information about total amount of the money to be paid, digital certificate of the customer, digital certificate of the merchant, Token ID, timestamp.

The first slot in the token contains information about the amount of money that has to be paid to the Merchant. The customer passes on this information to the Customer Bank in the order information OI. The Customer Bank will only release the amount of money mentioned in the token. The token also contains the digital certificate of the Customer and the digital certificate of the Merchant to verify that the token belongs to the particular Customer meant for the specific Merchant. There is also a Token ID which is unique to each transaction. The Token ID is a 256 bit code which is used once by the Customer Bank for every transaction. When the transaction for a particular Token ID is made, it is never generated again. A timestamp is also included in the token. The timestamp is included so that if any disputes arise, the arbitration body gets an authenticated proof of the date and time of the transaction. The structure of the Token is given in Figure 2:

| AM | DCert$_C$ | DCert$_M$ | TKN$_{ID}$ | TS |
|----|-----------|-----------|------------|-----|

Figure 2 Structure of the Token.

The Token ID in the token is encrypted with AES symmetric key with the Rijndael algorithm (Joan Daemen, Vincent Rijmen, 2002). The Customer Bank generates the symmetric key and decrypts it to check for any tampering in the Token when it is requested for the payment. The Customer Bank is obliged to pay to the Merchant's bank only that amount of money that is embedded in the token.

Moreover, the Customer's Bank keeps a duplicate copy of the Token every time the unique Token is generated. So, before releasing the transaction money, the Customer Bank compares the Token with the copy stored with it. If the Customer Bank finds any evidence of tampering in the Token, the transaction is stopped immediately. The Customer Bank then reports the TTP that the Token has been tampered with. The TTP then sends a message to the customer who requests the Customer Bank to generate a new Token and the whole process is carried out once again. So, the token ensures end to end security in the SET protocol, as any modification in the token will be immediately detected and the transaction process will be stopped.

The steps involved in the TSET protocol are:

1. The customer C browses the website of the merchant M and orders the goods.

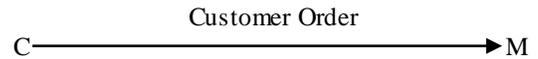

2. The merchant M sends his digital certificate DCert$_M$ along with the order information to the customer C to authenticate the merchant's validity.

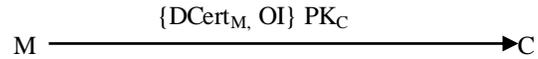

3. The customer logs into the TTP and checks the trust factor of the merchant. If the merchant is trustworthy of doing business, the customer proceeds to do the business otherwise abstains from it. If satisfied, the customer requests the Customer Bank CB for a token with the desired amount of money which the customer bank sends to the customer.

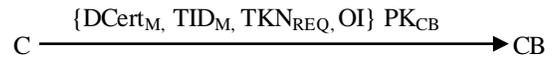

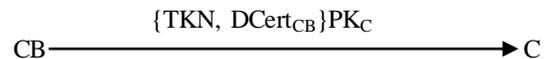

4. The customer C sends the purchase confirmation to the merchant by sending his digital certificate and Order Information to the merchant.

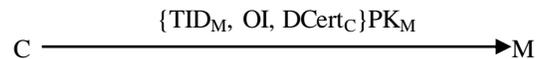

At the same time, the customer C sends the order information and token to the TTP.

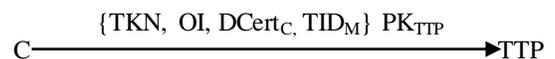

The transaction process of TSET model is shown in Figure 2

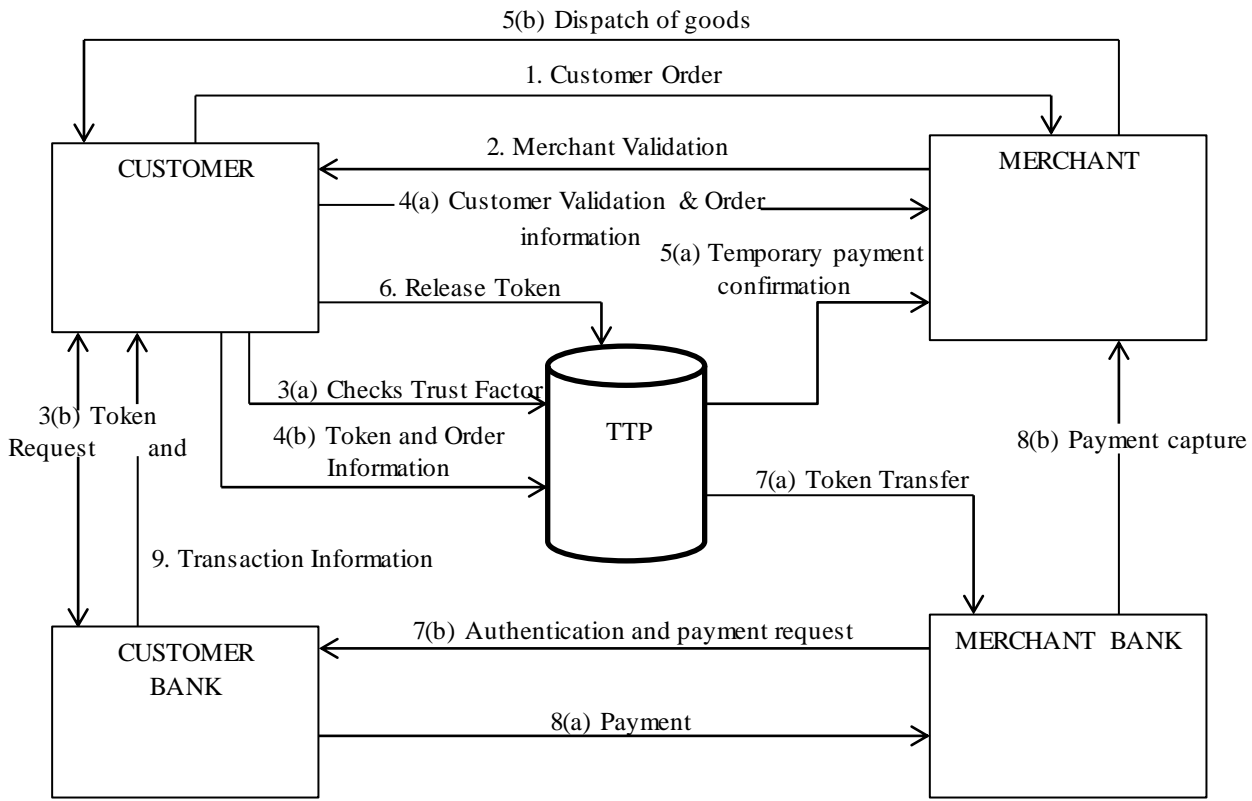

Figure2. The transaction process in the TSET protocol

5. The merchant decrypts the message from the customer and after the authentication sends the TTP request for confirmation of the temporary payment. The TTP meanwhile decrypts the Token and checks for authenticity. If the TTP acknowledges the receipt of temporary payment in the form of token, the merchant dispatches the goods.

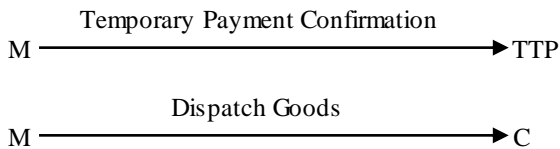

6. The customer after receiving the goods, if satisfied, sends a message to the TTP to release the token to the Merchant Bank MB. Otherwise, the customer asks the TTP to inform the merchant to replace the goods and hold the token for more time. In this case the trust factor of the merchant decreases.

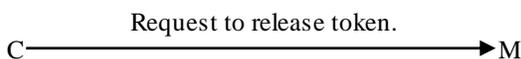

7. The merchant bank MB upon receiving the token sends the token to the customer bank and request customer bank for payment.

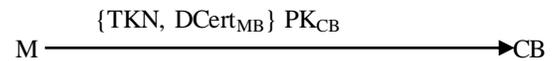

8. The customer bank on receiving the token decrypts the Token with its private key. The customer bank then decrypts the Token ID with its symmetric key and matches all the data in the Token against the data of the Token ID stored in its own database. It looks for any tampering in the token. If there is any tampering in the token, the CB reports it to the TTP that the token has been hacked. In this case, the TTP asks the customer for generation of a new token. If the token has not been tampered with, the customer bank CB sends the money to the merchant bank.

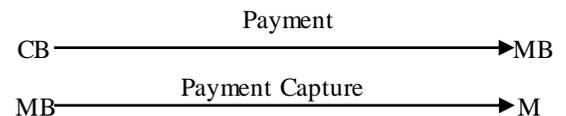

9. The merchant captures the payment from the Merchant Bank on completion of the transaction and notifies the TTP about it. On the other hand, the Customer Bank CB after the final transaction informs the customer that the transaction has been completed.

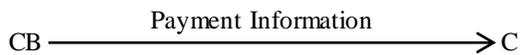

CB —— Payment Information ——> C

## V. ANALYSIS OF THE NEW PROTOCOL

This section discusses the various features of the TSET protocol.

### A. Trust Mechanism

Based on the new TSET protocol the customers can now have a prior knowledge of the trustworthiness of the merchants. It is up to the customer if he wishes to continue his transactions with the merchant having a certain trust value. This mechanism also ensures that the merchants will only provide the authentic products as desired by the customer otherwise their trust factor falls which has an impact on their business.

### B. Quality of Service

The new protocol entitles the customer to return the goods if he is not satisfied with it. The TTP acts as the governing body and ensures that the merchant provides a replacement or refund of the product within a stipulated period of time. Failing to resolve the discrepancy leads to decrease of trust factor of the merchant. This will not be desired by the merchant. So this protocol ensures that the merchant provides only products of the highest quality as desired by the customers.

### C. End to End Security

The new SET protocol ensures end to end security. At no point in time of the transaction can the Token, where the amount of transaction is embedded can be compromised by any agency. If the Token is altered and the amount embedded in the Token is tampered, the Customer Bank detects it by matching it with copy of token in its database. Evidence of any tampering will immediately result in halting of the transaction process and a new token will be generated. So, this ensures that the merchant will only get the desired amount of money as provided by the customer.

### D. Disputes

As every transaction has to pass through the TTP, it ensures a fair trading between all the parties. If there arises any dispute regarding the transactions, the TTP can provide the transaction log between the two parties. This record cannot be denied by either party and thus there can be settlement based on this record.

### E. Privacy

The protocol also partially fulfills the security requirements as mentioned in (J.J Hwang et al., 2003). The customer information is not known to the merchant or the merchant bank at any point of the transaction. Moreover, the order information is known only to the customer and merchant.

## VI. CONCLUSION

In this paper the Token based Secure Electronic Transaction for mobile payments has been discussed. We primarily focused on the trust factor and end to end security of the protocol and quality of service. Depending on the different trust values assigned to the merchants, the customers can now be sure of the trustworthiness of the merchant before indulging in the transaction process. The end to end security mechanism ensures that a faulty transaction never takes place and only the actual amount of money is released to the merchant. Because of the grading mechanism, the merchant will always try to provide good quality products to the customers so that their trust factor remains high. We believe that by increasing the trust of customers, improving the security of TSET protocol and by providing better quality of service, more and more people will be open towards electronic commerce.